# Charge pumping in hBN-encapsulated graphene driven by surface acoustic waves

Dublin M. Nichols[1], Jameson G. Berg[2], Takashi Taniguchi[3], Kenji Watanabe[4], Pallavi Dhagat[5], Vikram V. Deshpande[2], Albrecht Jander[5], and Ethan D. Minot[1]

[1] Department of Physics, Oregon State University, Corvallis, Oregon 97331, USA
[2] Department of Physics and Astronomy, University of Utah, Salt Lake City, Utah 84112, USA
[3] Research Center for Materials Nanoarchitectonics, National Institute for Materials Science, 1-1 Namiki, Tsukuba 305-0044, Japan
[4] Research Center for Electronic and Optical Materials, National Institute for Materials Science, 1-1 Namiki, Tsukuba 305-0044, Japan
[5] School of Electrical Engineering and Computer Science, Oregon State University, Corvallis, Oregon 97331, USA

**Abstract**

Surface acoustic waves (SAWs) on piezoelectric insulators can generate dynamic periodic potentials inside one-dimensional and two-dimensional materials. These periodic potentials have been utilized or proposed for various applications, including acoustoelectric charge pumping. In this study, we investigate acoustoelectric charge pumping in graphene with very low electrostatic disorder. By employing a graphite top gate on boron-nitride-encapsulated graphene, we adjust the graphene carrier concentration over a broad range, enabling us to examine the acoustoelectric signal in both mixed-carrier and single-carrier regimes. We discuss the benefits of hBN-encapsulated graphene for charge pumping applications and introduce a model that describes the acoustoelectric signal across all carrier concentrations, including at the charge neutrality point. This quantitative model will support future SAW-enabled explorations of phenomena in low-dimensional materials and guide the design of novel SAW sensors.

# 1 Introduction

Surface acoustic waves (SAWs) offer the possibility to create dynamic superlattices in 1D and 2D materials. When a SAW propagates across a strong piezoelectric insulator, the extension and compression of the insulator generates a periodic potential. SAWs can be generated with wavelengths ranging from tens of microns to tens of nanometers. In the burgeoning field of van der Waals (vdW) heterostructures made from 2D materials, SAWs have emerged as a new way to interact with charge carriers. For example, previous work has demonstrated the transport of indirect excitons in 2D semiconductor heterostructures [1], and contactless probing of quantum oscillations in graphene [2].

Interest in applying SAWs to 1D and 2D materials is inspired by previous experiments on GaAs/AlGaAs quantum wells, and by a number of outstanding theoretical proposals. For example, photogenerated electron-hole pairs in GaAs were separated by a SAW potential, and then released to generate photons [3]. SAWs were utilized in conjunction with Coulomb blockade to sequentially transport single electrons through a quantum point contact [4]. New insights into the quantum Hall effect and fractional quantum Hall effect in GaAs/AlGaAs quantum wells were obtained by utilizing commensurability effects with a SAW superlattice [5,6]. Turning to examples of theoretical proposals, Barnes et al. formulated a scheme for quantum computing using single electrons trapped in SAW potential minima ("flying qubits") [7–9]. Andreev recently proposed that a SAW applied to charge-neutral graphene can efficiently pump heat (approaching the Carnot limit) [10]. Talyanskii et al. proposed a scheme in which a SAW applied to a carbon nanotube can realize a topologically protected electron pump, defining a quantum standard for the current [11].

It is challenging to cleanly integrate SAWs with 1D and 2D electronic systems because many insulating surfaces (including piezoelectric insulators) introduce significant electrostatic disorder to the electronic system. This disorder disrupts the intrinsic properties of the low-dimensional material. To overcome this issue, Dean et al. introduced a method of encapsulating low-dimensional materials with hexagonal boron nitride (hBN), an ultraclean 2D insulator [12]. Boron-nitride-encapsulated graphene has been utilized for many applications; for example, demonstrating the highest-performing Hall-effect sensor [13]. However, there are no previous studies of acoustoelectric charge pumping in hBN-encapsulated graphene. In this work, we address the need for a detailed, quantitative analysis of the interaction between SAWs and graphene when the electrostatic disorder in the graphene is significantly reduced. Moreover, by using hBN encapsulation, we expect improvements such as larger pumping currents and increased sensitivity of pumping current with respect to carrier concentration, which may aid future technologies.

Previous authors have demonstrated acoustoelectric charge pumping in lower-quality graphene samples (see review by Hernández-Mínguez et al. [14]), but device quality (and sometimes non-tunable carrier concentration) has hindered quantitative comparison with theory. In our device design, the electric field from the SAW couples to the graphene from below, while a graphite top gate enables us to tune the carrier concentration in graphene over a wide range. We demonstrate that the SAW can drive extremely high 2D current density in hBN-encapsulated graphene when the system is tuned close to the charge neutrality point (CNP). At the largest SAW powers, we see signs of nonlinear effects, suggesting that the SAW causes a perturbation in carrier density that is comparable to the equilibrium carrier density. We present a theoretical model to describe the acoustoelectric current/voltage as a function of charge carrier concentration. In contrast to previous work, we extend the classical relaxation model to account for the coexistence of electrons and holes near the CNP. Our mixed-carrier model describes the observed acoustoelectric transport signals at all carrier concentrations, including the CNP.

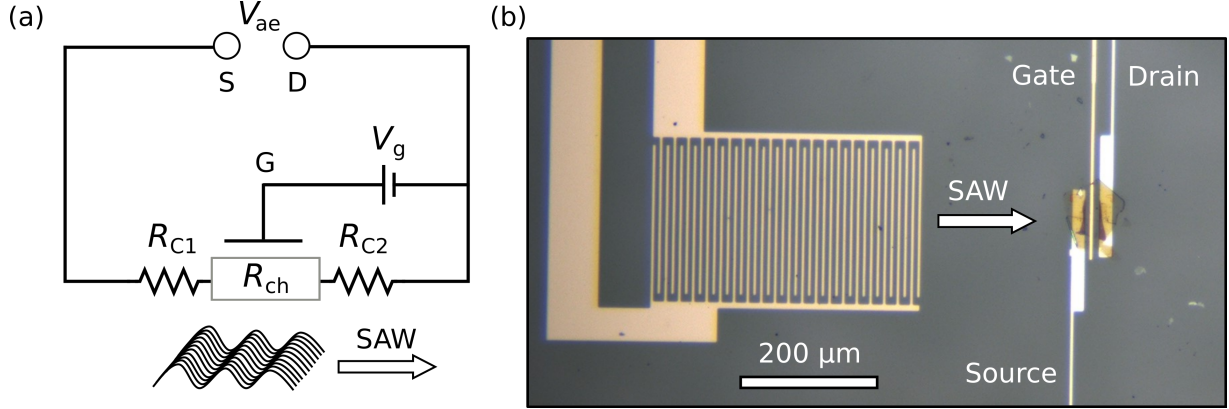

**Figure 1.** Overview of the experiment. (a) Optical microscope image of the interdigitated transducer (IDT) and an hBN-encapsulated graphene device (Device 1). The three electrodes are labeled source, drain, and gate. (b) The acoustoelectric voltage ($V_{ae}$), is measured across the source (S), drain (D) electrodes. The gate electrode (G) is used to tune carrier concentration. The total resistance of the graphene device includes the left and right contact resistance ($R_{C1}$ and $R_{C2}$) and the graphene channel resistance ($R_{ch}$).

## 2 Background

When the travelling electric field generated by a surface acoustic wave (SAW) passes over a conductive 2D material, charge carriers move in response to the electric field. This interaction transfers energy and momentum to the carriers and attenuates the SAW. The classical relaxation model predicts the attenuation of the SAW will be described by an attenuation constant [15]

$$\Gamma = K^2 \frac{\pi}{\lambda} \left[ \frac{(\sigma/\sigma_c)}{1 + (\sigma/\sigma_c)^2} \right], \qquad 1$$

where $K^2$ is the piezoelectric coupling constant, $\lambda$ is the wavelength of the SAW, $\sigma$ is the conductivity of the 2D material, and $\sigma_c = v_{SAW}(\epsilon_1 + \epsilon_2)$ is a characteristic conductivity defined by the properties of the substrate, where $\epsilon_1$ and $\epsilon_2$ are the dielectric permittivities above and below the 2D material and $v_{SAW}$ is the SAW velocity in the piezoelectric substrate. For a 2D material on LiNbO$_3$, $K^2 = 0.05$ and $\sigma_c \approx (1\ M\Omega)^{-1}$ [15,16].

If the 2D material is wired in a short-circuit configuration, the classical relaxation model predicts that the SAW will drive a net flow of charge through the 2D material [16]. The predicted short-circuit acoustoelectric current density is given by

$$j_{ae} = \pm \frac{\mu I}{v_{SAW}} \Gamma, \qquad 2$$

where the sign is determined by the sign of the carriers (negative for electrons, positive for holes), $\mu$ is the carrier mobility, and $I \propto P_{RF}/W$ is the SAW intensity at the 2D material, where $P_{RF}$ is the power applied to the interdigitated transducer (IDT) and $W$ is the aperture width of the IDT used to drive the SAW. In experiments, there is a decrease in power between the RF generator and the SAW (insertion loss) which can be described by a proportionality constant. Equations 1 and 2 assume a single carrier type (either electrons or holes) and predict that $j_{ae}$ is maximized when $\sigma = \sigma_c$. However, in graphene, the model must be adjusted to describe a system in the mixed-carrier regime.

## 3   Experiment Design

Figure 1 (b) illustrates our experimental measurement scheme. To detect the acoustoelectric voltage, $V_{ae}$, we probe the open-circuit voltage across the graphene channel. There is an effective force on charge carriers from the SAW which pushes charge carriers towards one side of the channel. Since there is no net current in the open-circuit configuration, an acoustoelectric voltage develops to balance the force from the SAW. The relationship between expected $V_{ae}$ (open-circuit voltage) and $j_{ae}w$ (short-circuit current) is $j_{ae}w = V_{ae}/R_{ch}$, where $R_{ch}$ is the resistance of the graphene channel and $w$ is the width of the graphene channel measured perpendicular to the SAW propagation direction. We use $V_{ae}$ for our analysis because a true measurement of short-circuit current requires zero-resistance contacts to the graphene, and a zero-impedance current amplifier.

Previous attempts to quantify $j_{ae}$ in graphene using the short-circuit current method likely suffered from the complication of large series resistance [20–23].

An IDT that emits a SAW with wavelength $\lambda = 20$ μm from an aperture $W = 230$ μm was fabricated on Y-cut black LiNbO$_3$ (University Wafer) using photolithography and metallization of 5/25 nm Cr/Au. Black LiNbO$_3$ was chosen because the material can tolerate faster thermal ramps than transparent LiNbO$_3$ [2,17]. Graphene devices were constructed next to the IDT as follows: First, we fabricated source/drain contacts (2.5/15 nm Cr/Pd) with channel spacing $l = 20$ μm (equal to λ) which sit 200 μm from the IDT. These contacts were cleaned using an atomic force microscope (AFM) in contact mode with a force of 100 nN [18]. Then, we exfoliated the few-layer graphene and hexagonal boron nitride (hBN) crystal flakes onto blank silicon wafers (300 nm oxide). We found large, uniform-thickness flakes using optical microscopy, and measured flake thicknesses using AFM. We used a strongly adhesive polycaprolactone (PCL) stamp to remove unwanted graphene flakes (reducing the likelihood that unwanted graphene flakes would short the IDT), and to tear the channel graphene flakes into rectangular pieces of a single thickness [19]. We then used a PC/PDMS (poly(bisphenol A carbonate)/polydimethylsiloxane) stamp and the standard dry transfer technique [20] to create the van der Waals heterostructure and to place the stack on the prefabricated source/drain contacts. Metal contact to the graphite top gate was made by a final photolithography and metallization step (2.5/45 nm Cr/Au).

Our aim was to create devices with different levels of electrostatic disorder so we could study the effect of disorder on the acoustoelectric signal. The graphene channel of Device 1 is fully encapsulated in hBN to minimize electrostatic disorder (see inset of Fig. 2c) [12]. The graphene channel of Device 2 is half-encapsulated such that the graphene lies directly on the LiNbO$_3$ substrate (see inset of Fig. 2d). The LiNbO$_3$ substrate induces electrostatic disorder in the graphene channel. For Device 1, we selected two hBN flakes, one graphene flake for the channel, and one graphite flake for the gate. The bottom hBN flake was selected to cover the gap between the source and drain contacts. The graphene flake was selected to extend beyond the bottom hBN flake and make electrical contact to the source and drain, as shown in Fig. 2c. The gate insulator of Device 1 is 29-nm hBN, giving gate capacitance $C_g = 0.11$ µF/cm$^2$. The gate insulator of Device 2 is 13-nm hBN, giving $C_g = 0.24$ µF/cm$^2$.

## 4 Results

Figure 2 (a) and (b) show $V_{ae}$ as a function of the spatially averaged carrier concentration in Device 1 (fully encapsulated) and Device 2. The transport curves have been shifted along the $V_g$-axis to align the CNP with $V_g = 0$. The open-circuit voltage was measured while driving the IDT at its resonance of 170 MHz using an Agilent N5183A analog signal generator, using a Stanford SR560 voltage preamplifier, filtered with the internal 30 Hz low-pass filter of the SR560. We verified that the frequency dependence of $V_{ae}$ closely follows the spectrally resolved measurement of RF power absorbed by the interdigitated transducer (see Supplementary Material). Figure 2 (a) and (b) show other features that are expected for acoustoelectric transport. At the CNP, the sign of $V_{ae}$ reverses (indicating that the SAW pushes electrons and holes in the same direction). The magnitude of $V_{ae}$ changes non-monotonically with increasing carrier density. At large carrier concentrations, $V_{ae}$ decays asymptotically toward zero, consistent with Eqs. 1 and 2.

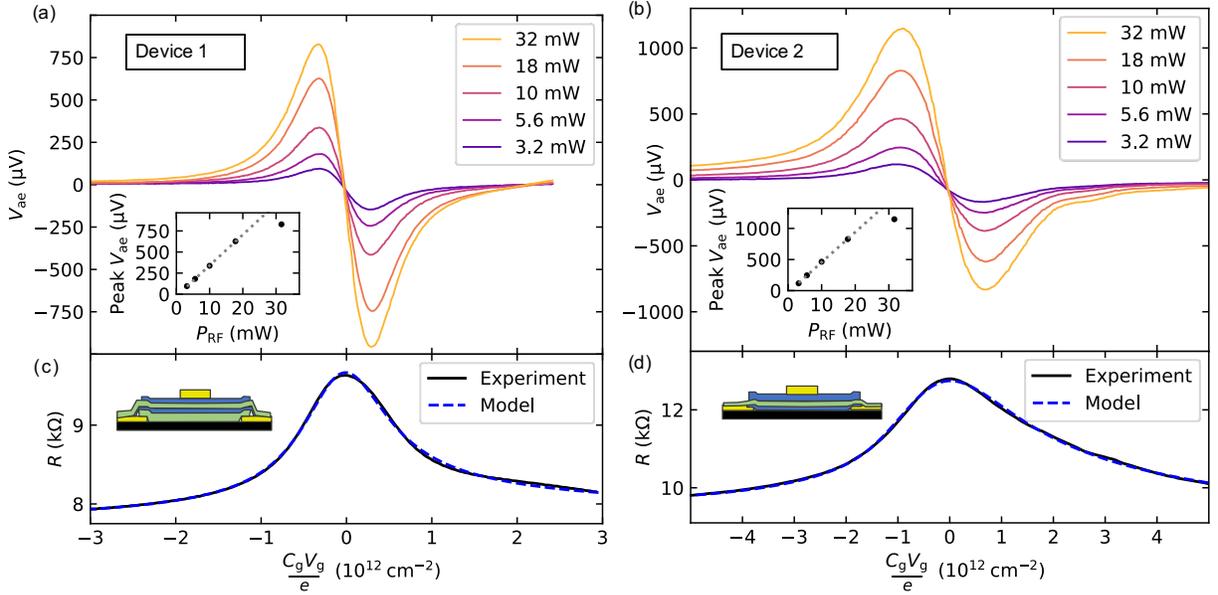

**Figure 2.** Room-temperature transport characteristics of Device 1 (fully encapsulated) and Device 2. (a)-(b) Acoustoelectric voltage ($V_{\text{ae}}$) as a function of the carrier concentration in the graphene channel, for various levels of applied RF power $P_{\text{RF}}$. $C_g$ is the gate capacitance per unit area (see Sec. 3). The inset shows the maximum $V_{\text{ae}}$ as a function of $P_{\text{RF}}$, with the grey dotted line as a guide for the eye. (c)-(d) Device resistance measured by setting $P_{\text{RF}} = 0$ and $V_{\text{SD}} = 100$ mV. The blue dashed line shows a fit to the experimental data. The insets show the device structures. Colors indicate LiNbO$_3$ (black), Pd (yellow), few-layer graphene (blue), and hBN (green).

Figure 2 (c) and (d) show DC transport in Devices 1 and 2, where $R = R_{\text{ch}} + R_{\text{C1}} + R_{\text{C2}}$. With the SAW power turned off, we applied a source-drain bias $V_{sd} = 100$ mV and measured the DC current using a Stanford SR570 current preamplifier while sweeping $V_g$. As expected, the device resistance is largest at the CNP, where the number of free carriers is minimized. The device resistance asymptotically approaches a finite value at large $V_g$. We associate this finite resistance with contact resistance $R_{\text{C1}} + R_{\text{C2}}$.

The insets of Figure 2 (a) and (b) show the maximum value of $V_{ae}$ versus applied SAW power $P_{RF}$. In both devices, we observe a linear relationship between $P_{RF}$ and $V_{ae}$ up to 18 mW, which is consistent with Eq. 2 and has been observed in previous acoustoelectric graphene devices [21]. However, at $P_{RF} = 31$ mW, the acoustoelectric voltage is smaller than expected, suggesting a nonlinear relationship between $P_{RF}$ and $V_{ae}$. Non-linear behavior has been observed previously in GaAs acoustoelectric devices at high SAW powers [22], and suggests that the perturbation of carrier density due to the SAW is comparable in magnitude to the average carrier density in the graphene. More work is needed to understand the origin of this nonlinear relationship in acoustoelectric graphene devices. Therefore, for the analysis below, we will use the $P_{RF} = 18$ mW data to ensure that the linear model is applicable.

Of principal interest to this work is the shape of the peak in $V_{ae}$ seen in Figure 2 (a) and (b). The single-carrier, classical relaxation theory (Eq. 1 and 2) predicts that the acoustoelectric signal should reach a maximum when the channel conductivity $\sigma$ is equal to the characteristic conductivity of the piezoelectric substrate. For LiNbO$_3$, this characteristic conductivity is $\sigma_c \approx (1\ \text{M}\Omega)^{-1}$ [15,16]. However, as can be seen in Figure 2a, the peak in $V_{ae}$ occurs when $R_{ch} \approx 2$ k$\Omega$, which corresponds to $\sigma \approx (2\ \text{k}\Omega)^{-1}$. Prior authors noted similar discrepancies between the single-carrier theory and graphene measurements, but no quantitative description for the peak has been previously reported in literature. The peak shape can be accurately described using the mixed-carrier model outlined below.

## 5 Discussion

First, we compare the magnitude of the acoustoelectric signal measured in our hBN-encapsulated graphene device to the results of prior graphene-based acoustoelectric experiments. For Device 1, $w = 15$ μm, $R_{ch} \approx 2$ k$\Omega$, and maximum $|V_{ae}| = 0.98$ mV. If the channel was measured in a short-circuit configuration (assuming we decreased the contact resistance, such that $R_{C1} + R_{C2} \ll R_{ch}$), we would expect $j_{ae} \approx 33$ mA/m. This current density is nearly 10 times higher than previous measurements of acoustoelectric current in graphene devices [23–26]. For Device 2, we infer $j_{ae} = 18$ mA/m. The higher current density obtainable in the fully encapsulated graphene is evidence that lowering electrostatic disorder boosts acoustoelectric signals. We further quantify this insight with the model presented below.

To understand the gate-voltage dependence of $V_{ae}$ in graphene devices, we consider the co-existence of electrons and holes at the CNP. In electrostatically gated graphene that has no disorder and no thermally activated charge carriers, we expect the electron and hole concentrations to be $n = C_g V_g$ for $V_g > 0$, and $p = C_g |V_g|$ for $V_g < 0$, where $C_g$ is the gate capacitance per unit area (illustrated in Fig. 3c). In a real graphene sample, there are thermally activated carriers and spatial fluctuations in electrostatic potential that modify the electron and hole concentrations. Figure 3a illustrates the spatial inhomogeneity of carrier concentration in graphene [27]. To model this, we assume a position-dependent gate voltage offset $V_{\text{offset}}(x, y)$ which has an average value 0 and standard deviation $\delta V$. We assume that the distribution of $V_{\text{offset}}$ values follow a normal distribution. A similar disorder model has been used to describe the gate-dependent Hall effect in graphene [28]. From the function $V_{\text{offset}}(x, y)$, we obtain integral forms for the spatially averaged carrier concentrations in the graphene sample (see Supplementary Material) which can be solved analytically, giving

$$n(V_g, \delta V) = \frac{C_g}{e}\left(\frac{V_g}{2}\left(1 + \text{erf}\left(\frac{V_g}{\sqrt{2}\delta V}\right)\right) + \frac{\delta V}{\sqrt{2\pi}}\exp\left(-\frac{V_g^2}{2\delta V^2}\right)\right), \qquad 3$$

$$p(V_g, \delta V) = -\frac{C_g}{e}\left(\frac{V_g}{2}\left(1 - \text{erf}\left(\frac{V_g}{\sqrt{2}\,\delta V}\right)\right) - \frac{\delta V}{\sqrt{2\pi}}\exp\left(-\frac{V_g^2}{2\delta V^2}\right)\right), \qquad 4$$

where erf is the error function, and $e$ is the electron charge. Equations 3 and 4 are plotted in Fig. 3c. In this model, the minimum carrier concentration in the graphene (the sum of $n$ and $p$ at $V_g = 0$) is

$$(n_0 + p_0) = \sqrt{\frac{2}{\pi}}\frac{C_g \delta V}{e}. \qquad 5$$

The mixed-carrier model (Eq. 3 and 4) yields an accurate fit to the DC transport data, shown in Figure 2 (c) and (d). The dashed lines in Fig. 2(c) and (d) are constructed by assuming $\sigma = e\mu(n + p)$, and $R_{ch} = l/(w\sigma)$, where $l$ is the length of the graphene channel (20 μm for both devices). For Device 1 we find $\mu = 7150\ \text{cm}^2/(\text{Vs})$ and $(n_0 + p_0) = 0.38 \times 10^{12}\ \text{cm}^{-2}$. The minimum carrier concentration in Device 1 is approaching the room-temperature limit ($\approx 0.16 \times 10^{12}\ \text{cm}^{-2}$) which corresponds to the concentration of thermally activated charge carriers in ultraclean charge-neutral graphene [30]. For Device 2 we find $\mu = 3500\ \text{cm}^2/(\text{Vs})$ and $(n_0 + p_0) = 0.97 \times 10^{12}\ \text{cm}^{-2}$. The fully encapsulated device (Device 1) has higher mobility and significantly reduced charge disorder compared to Device 2.

To describe acoustoelectric voltage, we combine the mixed-carrier model (Eq. 3 and 4) with the classical relaxation model (Eq. 1 and 2). The SAW pushes both electrons and holes in the same direction (Fig. 3b). The fraction of these moving carriers which are uncompensated (carriers which do not have a partner of opposite sign) is given by $(p - n)/(p + n)$. Using this fraction, we modify Eqs. 1 and 2, finding the net current density from electrons and holes to be

$$j_{ae} = \pm \frac{\mu I}{v_{SAW}} K^2 \frac{\pi}{\lambda}\left[\frac{(e\mu(n+p)/\sigma_c)}{1 + (e\mu(n+p)/\sigma_c)^2}\right]\left(\frac{p-n}{p+n}\right). \qquad 6$$

In our experiment, $e\mu(n+p) \geq 500\,\sigma_c$ at all gate voltages, therefore Eq. 6 simplifies to

$$j_{ae} \approx \pm \frac{I\sigma_c}{ev_{SAW}} K^2 \frac{\pi}{\lambda}\left(\frac{p-n}{(p+n)^2}\right). \qquad 7$$

At large $V_g$, where $n \gg p$ (or $p \gg n$), the model predicts $j_{ae} \propto 1/V_g$, consistent with the single carrier model (Eq. 2). Equation 7 can be extended to acoustoelectric voltage using the relationship $V_{ae} = j_{ae}wR_{ch}$. At large $V_g$, the model predicts $V_{ae} \propto 1/V_g^2$. Unlike the single carrier model, Eq. 7 is also valid at small $V_g$ where electrons and holes coexist.

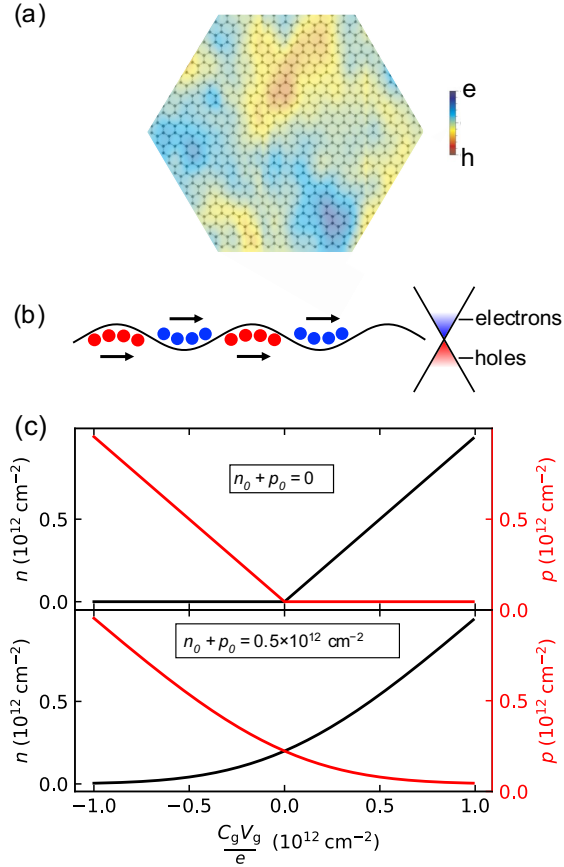

**Figure 3.** (a) A spatial map of electron and hole inhomogeneity in graphene (adapted from Ref. [27] with permission). (b) At the CNP, there are equal populations of electrons and holes. Electrons and holes are pushed in the same direction by the SAW, so there is no net acoustoelectric current. (c) Top: The electron concentration (black) and hole concentration

(red) in a graphene device with no thermally activated charge carriers and no spatial fluctuation in electrostatic potential. Bottom: The electron and hole concentrations when there is a spatial fluctuation in electrostatic potential.

Figure 4 shows the excellent fit between our mixed-carrier model prediction and the measured $V_{ae}$ curves. The key fitting parameter, $\delta V$, controls the width of the $V_{ae}$ peaks. For Device 1 (fully encapsulated), the best fit yields $\delta V = 0.65$ V, which is equivalent to $(n_0 + p_0) = 0.35 \times 10^{12}$ cm$^{-2}$. For Device 2, the disorder parameter is more than doubled, $(n_0 + p_0) = 0.88 \times 10^{12}$ cm$^{-2}$.

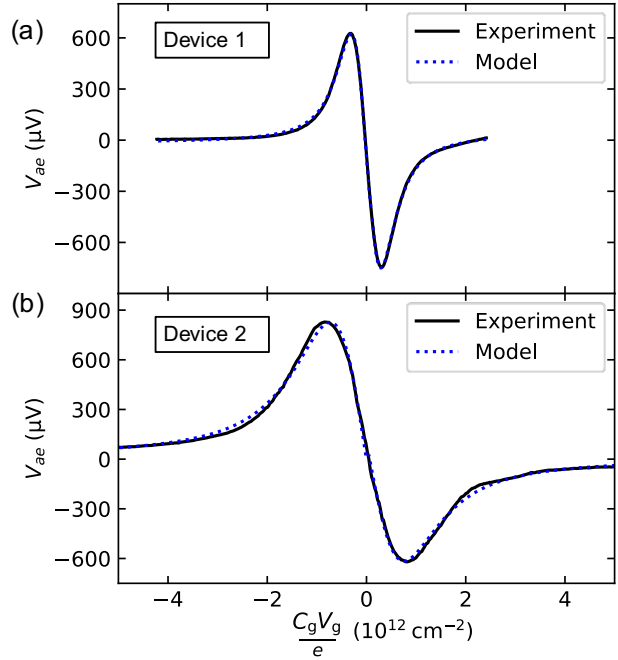

**Figure 4.** Mixed-carrier model (Eq. 7) fit to acoustoelectric transport in the fully encapsulated (a) and half-encapsulated (b) graphene devices. This data is taken with $P_{RF} = 18$ mW.

The height of the $V_{ae}$ peak for electron-doping differs slightly from the height of the $V_{ae}$ peak for hole-doping. Therefore, we used separate fitting parameters for peak height on either side of $V_g = 0$. A similar asymmetry is found in gate-dependent Hall-effect measurements of graphene and the mechanism has been discussed extensively in literature [31–33]. Using our three-parameter fit ($\delta V$ plus two peak height parameters) we achieve excellent quantitative agreement with the experimental data.

Our mixed-carrier model (Eq. 6) gives a satisfying explanation for why $\sigma \neq \sigma_c$ when $V_{ae}$ is maximized in a graphene device. To maximize $V_{ae}$, the gate voltage must be tuned away from $V_g = 0$ to avoid the cancellation of electron current by hole current. Conversely, if $V_g$ is too large, then $V_{ae}$ decays as $1/V_g^2$. Thus, there is a sweet spot in gate voltage where most carriers have the same polarity, but carrier concentration is still small. The position of this sweet spot is determined by the disorder parameter $\delta V$. Reducing $\delta V$ will increase the height of the $V_{ae}$ peak.

The high sensitivity of $V_{ae}$ with respect to $V_g$ at the CNP ($V_g = 0$) suggests that charge pumping in ultraclean graphene may be useful for sensing applications. For example, adsorption of gas molecules onto a graphene surface can modulate $\sigma$ by modulating the concentration of charge carriers in the graphene. Prior work has confirmed that changes in $\sigma$ can be used to detect adsorbed gas (reviewed in Ref. [34]). However, the $\sigma$-based transduction mechanism does not work at the CNP where $d\sigma/dV_g = 0$. In contrast, the acoustoelectric voltage, $V_{ae}$, is most sensitive to changes in $n$ and $p$ when the device is operated at the CNP ($dV_{ae}/dV_g$ is maximal). Working at the CNP, gas detection events would correspond to an increase or decrease in $V_{ae}$ from zero (a small signal on top of zero background), which is preferable to detecting a small change in $\sigma$ on top of a large background. Additionally, open-circuit voltage measurements circumvent unwanted noise that is generated by fluctuating contact resistance [35]. Further work in this direction could be pursued by modifying the architecture of Device 1: the top side of graphene could be exposed to the environment, while the bottom side of graphene would rest on hBN, which in turn would rest on LiNbO$_3$.

## 6  Conclusions

We have demonstrated that the acoustoelectric signals in graphene (voltage and current) can be significantly increased by minimizing charge disorder in the graphene. hBN-encapsulated graphene allows us to reach lower carrier concentration (close to the thermal limit) so that channel resistance better matches the optimal value to absorb SAW power. Our measurements demonstrate that room-temperature acoustoelectric current density in graphene can reach at least 33 mA/m (nearly ten times larger than previous reports). We have presented a quantitative model for the gate-dependent acoustoelectric signals that describes the coexistence regime where both electrons and holes are present. This quantitative framework will aid future SAW-based experiments designed to probe new phenomena in graphene and other 2D materials.

**Supplementary Material**

Supplementary material contains optical images of Device 1 and Device 2, integral forms of Eqs. 3 and 4, spectrally resolved measurements of reflected SAW power and acoustoelectric current, and comparison of pumped current density in prior gated acoustoelectric graphene devices.

**Acknowledgements**

This work was supported by the National Science Foundation under Grant No. 2004968. Part of this research was conducted at the Northwest Nanotechnology Infrastructure, a National Nanotechnology Coordinated Infrastructure site at Oregon State University which is supported in part by the National Science Foundation (grant NNCI-2025489) and Oregon State University.

K.W. and T.T. acknowledge support from the JSPS KAKENHI (Grant Numbers 21H05233 and 23H02052) and World Premier International Research Center Initiative (WPI), MEXT, Japan.

**Conflict of Interest Statement**

The authors have no conflicts to disclose.

**Data Availability Statement**

The data that support the findings of this study are available from the corresponding author upon reasonable request.

**Supplementary Material**

**S1 Device images**

Figure S1 shows optical microscope images of Device 1 (fully encapsulated) and Device 2. Using tapping-mode atomic force microscopy (AFM), we measured the thickness of the channel graphene flakes to be 1.2 nm for Device 1 and 1.5 nm for Device 2.

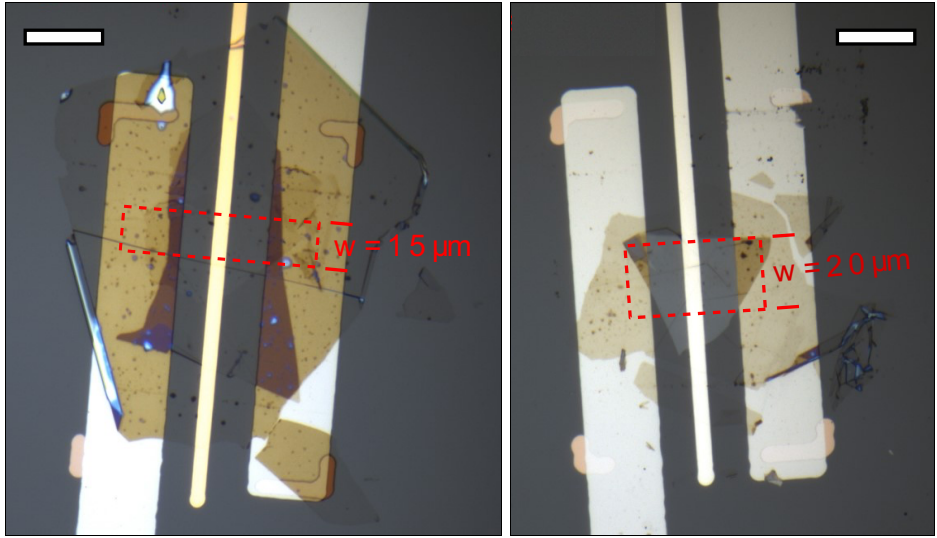

**Figure S1.** Optical images of Device 1 (left) and Device 2 (right). Due to poor contrast of graphene on LiNbO$_3$, we determined the approximate final location of the graphene channels from optical images taken during the transfer process. The approximate location and shape of the graphene channels are indicated with a red dotted line. Scale bar = 20 $\mu m$.

**S2 Spatially averaged carrier concentrations**

The probability of an *x-y* position in the graphene channel having a voltage offset $V_{\text{offset}}(x, y)$ is given by the probability distribution function

$$P(V_{\text{offset}}, \delta V) = \frac{1}{\sqrt{2\pi} \cdot \delta V} \exp(-V_{\text{offset}}^2/(2\delta V)^2). \qquad \text{S1}$$

Using this probability distribution function, we calculate the spatially averaged carrier concentrations in the graphene channel when the global gate is set to $V_g$

$$n(V_g) = \frac{C_g}{e} \int_{-V_g}^{\infty} P(V_{\text{offset}}, \delta V)(V_g + V_{\text{offset}}) dV_{\text{offset}}, \qquad \text{S2}$$

$$p(V_g) = \left| \frac{C_g}{e} \int_{-\infty}^{-V_g} P(V_{\text{offset}}, \delta V)(V_g + V_{\text{offset}}) dV_{\text{offset}} \right| \qquad \text{S3}$$

where $C_g$ is the gate capacitance per unit area. The lower limit of integration in Eq. S2 corresponds to the lowest value of $V_{\text{offset}}$ that still gives electron doping (see Fig. S2). Similarly, the upper limit of integration in Eq. S3 corresponds to the highest value of $V_{\text{offset}}$ that still gives hole doping.

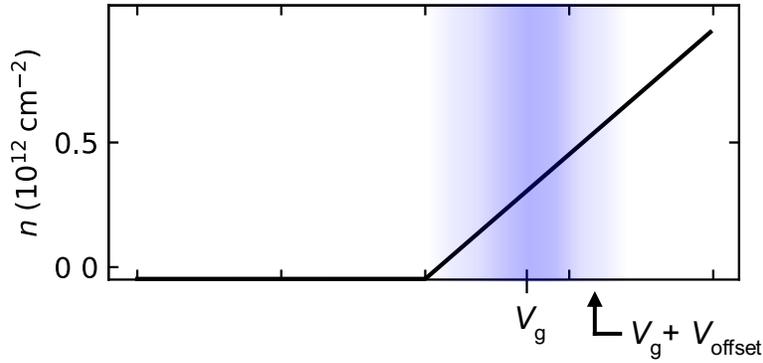

**Figure S2.** Calculating the spatially averaged electron concentration. The average gate voltage is $V_g$, and the local gate voltage is $V_g + V_{offset}$. The distribution of local gate voltages (depicted with blue shading) has a standard deviation $\delta V$. When $V_g + V_{offset} < 0$, the location does not contribute to electron concentration.

**S3 SAW device characterization and measurement of acoustoelectric current**

Figure S3 shows acoustoelectric current ($I_{ae}$) and reflected SAW power $S_{11}$ as a function of frequency in Device 2, measured using an Agilent E5071C-280 vector network analyzer. We observe a SAW resonance in $S_{11}$ at 170 MHz, close to our designed frequency of 174 MHz. $I_{ae}$ closely follows the drop in $S_{11}$, confirming the dependence of our measured acoustoelectric signals ($I_{ae}$ and $V_{ae}$) on transmitted SAW power. The IDTs in Device 1 and Device 2 are identical.

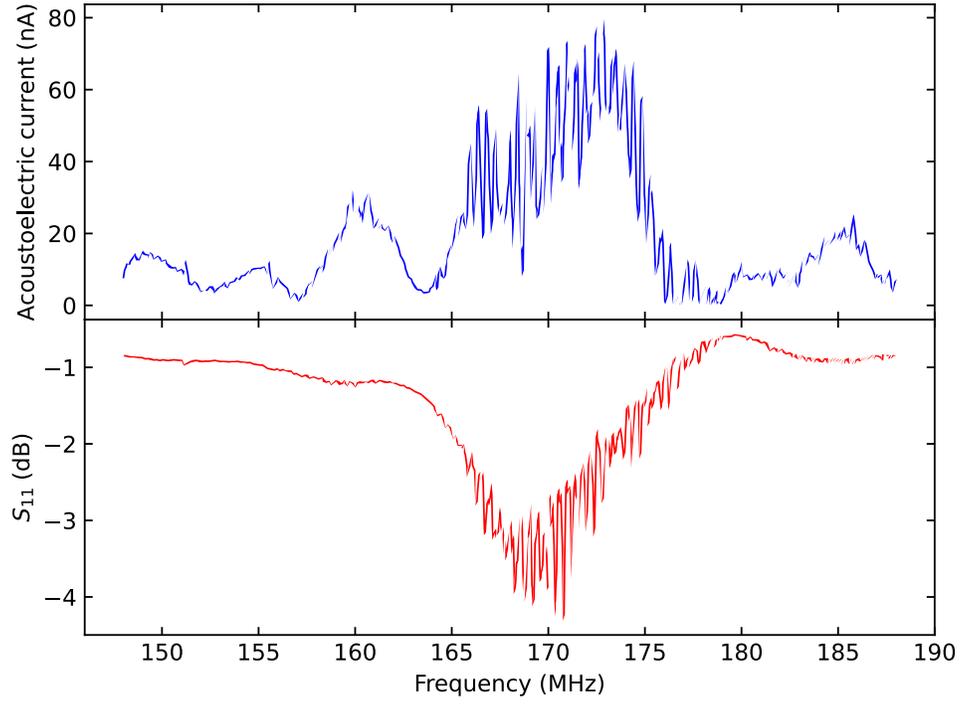

**Figure S3.** Acoustoelectric current (top) and reflected SAW power $S_{11}$ (bottom) as a function of frequency in Device 2. This data is taken with $P_{RF} = 10 \: mW$.

## S4 Comparing current density of prior gated acoustoelectric graphene devices

Ref. [25]: $I_{ae} = 17 \: \mu A, w = 5 \: mm: j = 3.4 \: mA/m$

Ref. [23]: $I_{ae} = 5 \: \mu A, w = 5 \: mm, j = 1 \: mA/m$

Ref. [24]: $I_{ae} = 0.1 \: \mu A, w = 80 \: \mu m: j = 1.25 \: mA/m$

Ref. [26]: $I_{ae} = 10 \: nA, w = 3 \: mm: j = 0.0033 \: mA/m$